\documentclass[twocolumn,showpacs,preprintnumbers,amsmath,amssymb]{revtex4}

\usepackage{graphicx}
\usepackage{dcolumn}
\usepackage{bm}

\begin{document}

\title{Better synchronizability predicted by a new coupling method}
\author{Ming Zhao$^1$}
\author{Tao Zhou$^1$}
\email{zhutou@ustc.edu}
\author{Bing-Hong Wang$^1$}
\author{Qing Ou$^2$}
\author{Jie Ren$^1$}
\affiliation{%
$^1$Department of Modern Physics, University of Science and
Technology of China, Hefei 230026, PR China\\
$^2$Department of Automation, University of Science and Technology
of China, Hefei 230026, PR China
}%

\date{\today}

\begin{abstract}
In this paper, inspired by the idea that different nodes should
play different roles in network synchronization, we bring forward
a coupling method where the coupling strength of each node depends
on its neighbors' degrees. Compared with the uniform coupled
method and the recently proposed Motter-Zhou-Kurths method, the
synchronizability of scale-free networks can be remarkably
enhanced by using the present coupled method, and the highest
network synchronizability is achieved at $\beta=1$ which is
similar to a method introduced in [AIP Conf. Proc. 776, 201
(2005)].
\end{abstract}

\pacs{89.75.Hc, 89.75.-k, 05.45.Xt, 87.18.Sn}

\maketitle

\section{Introduction}
Many collective dynamics in social, biological and communication
systems can be properly described by complex networks. These
networks exhibit complex topological properties such as the
small-world effects and the scale-free properties
\cite{Review1,Review2,Review3,Review4}. Many kind of network
models have been made to embody these properties. The so-called
small-world networks are the intermediates of regular lattices and
random networks in structure but bear both characters of the two
kind of networks, that is, they have small average distance as
random networks and large clustering coefficient as regular ones
\cite{WS}. The scale-free networks are a kind of small-world
networks with degree distribution obeying a power-law form. A
scale-free network can be created by successively adding new nodes
to the network and connecting them with the already existing ones
by the preferential attachment rule \cite{SFN}.

The interesting topological properties of complex networks make
the dynamics taking place on them much different from those on
regular or random ones. For example, coupled dynamical oscillators
on small-world networks are much easier to synchronize than on
regular lattices, and increasing the proportion of shortcuts of
networks will make the oscillators more synchronizable
\cite{Lago00,HuJK00,BJKim02,Zhou06}. It has also been observed
that the more heterogeneous of the network degree distribution is
the harder for the oscillators on the network to synchronize
\cite{LaiYC03}. Therefore, generally speaking, networks with short
average distance and homogeneous degree distribution will make the
oscillators on them more synchronizable
\cite{LaiYC03,Donetti05,Zhao06,Wangb05}.

Very recently, motivated by practical requirements and theoretical
interest, numbers of researches have begun to study how to enhance
the network synchronizability, especially for scale-free networks
\cite{Motter05,MotterAIP05,MotterEL05,Zhao05,Chavez05,Chavez06}.
The method proposed by Zhao \emph{et al.} \cite{Zhao05} can
sharply reduce the maximal betweenness thus enhance the network
synchronizability, but it will bring some economic and technologic
problems since the network structure is slightly changed. The
method proposed by Chavez \emph{et al.} keeps the network topology
unchanged, while adding some weight into the system
\cite{Chavez05,Chavez06}. However, to compute the weight, this
method needs the global structural information, which is usually
unavailable in huge communication systems. Therefore, in this
paper, we keep the network topology unchanged, and concentrate on
the coupling method using only the local information. The
Motter-Zhou-Kurths (MZK) method \cite{Motter05} is a typical
example, in which the coupling strength from a node $i$ is inverse
to its degree $k_i$. In MZK method, every neighbor of a node has
the same influence (coupling strength) to this node. However, in
real networks, different nodes may have different influences. For
example, in society, some people have strong influence on others
in some aspect but they are not influenced at the same level.
Another impressing phenomenon is that in the World Trade Web, the
small countries' economies fluctuate with the powerful countries
tightly, but the contrary does not occur \cite{LiX03}. Thus here,
based on the assumption that different nodes play different roles,
we adjust the influencing strength of each node receiving from
their neighbors according to the neighbors' degrees. That is, a
node is not influenced by its neighbors equally. It is found that
oscillators on scale-free networks coupled in this way can have
much stronger propensity for synchronization than in the previous
ways, with the exception of a recently introduced method
\cite{MotterAIP05,Nishikawa2006} , where the former is similar to
$\\beta =1$ in our model and is shown here to be the most
efficient method based on the information provided by the degree
of nearest neighbors only.

\begin{figure*}[!tp]
\centering
\includegraphics[width=\textwidth]{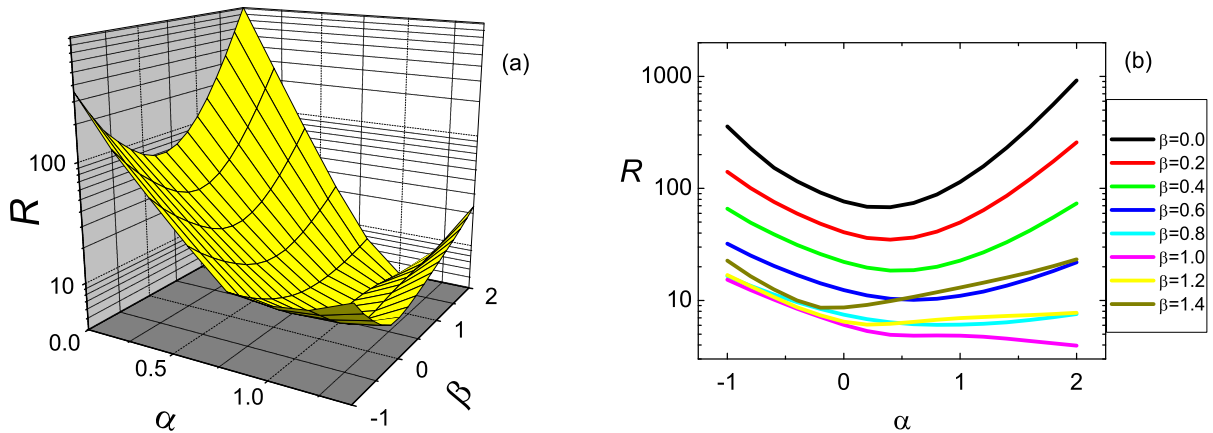}
\caption{(color online) (a) $R$ in the parameter plane ($\alpha$,
$\beta$). (b) $R$ vs $\alpha$ for different parameter $\beta$. The
numerical simulations are implemented based on the BA network of
size $N=1024$ and with average degree $\bar{k}=6$. The data are
obtained over 10 independent realizations.}
\end{figure*}

This paper is organized as follow: in section 2, the dynamical
equations of coupled oscillators and the master stability function
will be briefly introduced. In section 3, we will give the
simulation and analysis about synchronization of correlated
scale-free networks. Finally, we will draw our conclusion in
section 4.

\section{The dynamical equations and master stability function}

For a network of $N$ linear coupled identical oscillators, the
dynamical equation of each oscillator can be written as
\begin{equation}
\dot{\textbf{x}}^i=\textbf{F}(\textbf{x}^i)-\sigma\sum_{j=1}^NG_{ij}\textbf{H}(\textbf{x}^j),\hspace*{1em}i=1,2,...,N,
\end{equation}
where $\dot{\textbf{x}}^i=\textbf{F}(\textbf{x}^i)$ governs the
dynamics of individual oscillator, $\textbf{H}(\textbf{x}^j)$ the
output function, $\sigma$ the coupling strength, and $G_{ij}$ the
elements of the $N\times N$ coupling matrix. To guarantee the
synchronization manifold an invariant manifold, the matrix $G$
should has zero row-sum. Traditionally, the oscillators are
coupled symmetrically with uniform coupling strength and the
coupling matrix $G$ has the same form as Laplacian matrix $L$,
that is, $G_{ij}=L_{ij}$, where
\begin{equation}
    L_{ij}=\left\{
    \begin{array}{cc}
    k_i   &\mbox{for $i=j$}\\
     -1    &\mbox{for $j\in\Lambda_i$}   \\
     0    &\mbox{otherwise},
    \end{array}
    \right.
\end{equation}
where $k_i$ is the degree of node $i$ and $\Lambda_i$ is the set
of $i$'s neighbors. Because of the symmetry and the positive
semidefinite of $L$, all its eigenvalues are nonnegative reals and
the smallest eigenvalue $\lambda_0$ is always zero, for the rows
of $L$ have zero sum. And if the network is connected, there is
only one zero eigenvalue. Thus, the eigenvalues can be ranked as
$\lambda_0<\lambda_1\leq\lambda_2\leq...\leq\lambda_{N-1}$.
According to the criteria of master stability function
\cite{Pecora98,Pecora02,Wang2002a,Wang2002b}, the network
synchronizability can be measured by the eigenratio
$R=\lambda_{N-1}/\lambda_1$: The smaller it is the better the
network synchronizability and vice versa.

It is later found that networks with high heterogeneity of degree
distribution coupled uniformly are hard to synchronize. As
mentioned above, to eliminate this problem, Motter, Zhou and
Kurths suggested the coupling matrix taking the form
$G_{ij}=L_{ij}/k_i^\beta$ \cite{Motter05,MotterAIP05,MotterEL05}.
This simple change of coupling matrix enhances the network
synchronizability sharply, and the optimal condition is $\beta=1$.
And by exploiting the information contained in the load of each
edge (i.e. set the off-diagonal elements of the zero row-sum
coupling matrix $G$ to be
$G_{ij}=l_{ij}^\alpha/\sum_{j=1}^Nl_{ij}^\alpha$, where $l_{ij}$
is the load of edge connecting node $i$ and $j$), further
enhancement in synchronization is achieved
\cite{Chavez05,Chavez06}.

\begin{figure*}[!tp]
\centering
\includegraphics[width=\textwidth]{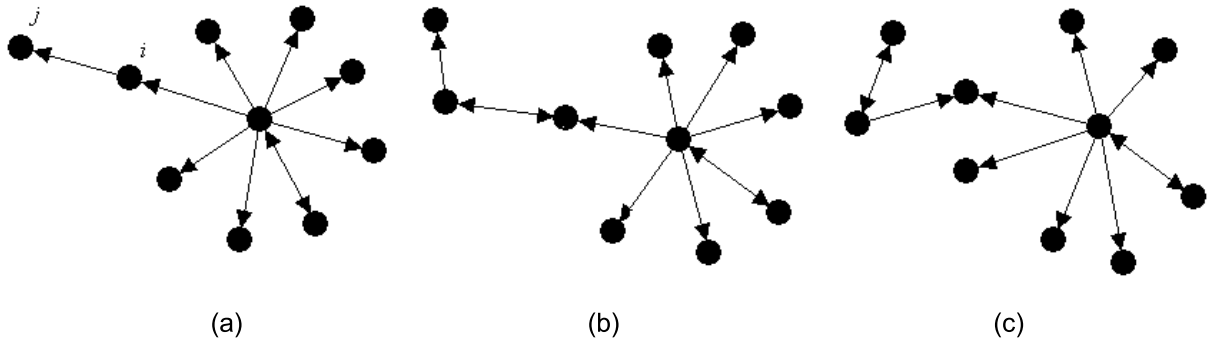}
\caption{The sketch maps of three simple equivalent networks,
where the arrow form node $i$ to node $j$ indicates the latter
receives coupling signal from the former. Their eigenratios are 2
(a), 6.8284 (b) and +$\infty$ (c), respectively.}
\end{figure*}

\begin{figure}
\scalebox{0.65}[0.65]{\includegraphics{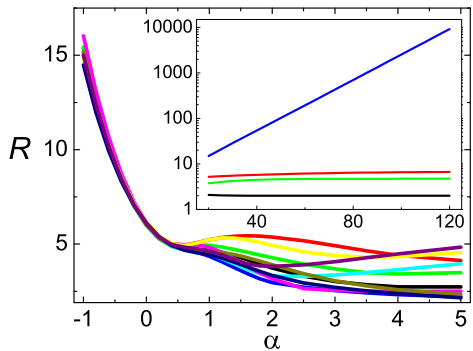}} \caption{(color
online) The eigenratio $R$ vs parameter $\alpha$ at $\beta=1.0$
for several BA network configurations of size $N=1024$ and with
average degree $\bar{k}=6$. Each color represents one
configuration.}
\end{figure}

Many real-world networks are highly heterogeneous with a few
nodes, named hubs, having very large degrees. When using the
uniform coupling method, these hubs synchronize first, and slowly
the nodes with fewer degree synchronize to them \cite{McGraw2005}.
If the influence of the hubs on the low-degree nodes becomes
stronger, the latter will synchronize to the former much easier,
obviously, the network synchronizability will be enhanced.
Therefore, we argue that not only reducing the communication load
of hubs (as did in MZK method), but also increase their influences
may further enhance the network synchronization.

Here we take into account the effects of different degrees of
nodes on synchronization, that is, a node in complex network is
not coupled uniformly by its neighbors but the coupling strength
is modulated by $k^\alpha$. Thus the coupling matrix $G$ takes the
form
\begin{equation}
    G_{ij}=\left\{
    \begin{array}{cc}
    S_i/S_i^\beta   &\mbox{for $i=j$}\\
     -k_j^\alpha/S_i^\beta    &\mbox{for $j\in\Lambda_i$}   \\
     0    &\mbox{otherwise},
    \end{array}
    \right.
\end{equation}
where $S_i=\sum_{j\in \Lambda_i}k_j^\alpha$. When
$\alpha=\beta=0$, this coupling scheme degenerates to the uniform
coupling scheme \cite{Pecora98}, the case of $\alpha=0$
corresponds to the MZK method \cite{Motter05}, and the case of
$\beta=1$ is equivalent to the one introduced in the ref.
\cite{MotterAIP05} (see the Eq.(15) for details).

Using a similar method to the one proposed by Motter \emph{et al}.
\cite{MotterEL05}, we next prove that all the eigenvalues of
matrix $G$ are real. Note that, Eq. (3) can be written as
\begin{equation}
G=DL',
\end{equation}
where
$D=\mbox{diag}\{k_1^{-\alpha}S_1^{-\beta},k_2^{-\alpha}S_2^{-\beta},...,k_N^{-\alpha}S_N^{-\beta}\}$
is a diagonal matrix, and $L'=(L_{ij}')$ is a symmetric zero
row-sum matrix, whose off-diagonal elements are
$L_{ij}'=k_i^\alpha k_j^\alpha$. From the identity
\begin{equation}
\mbox{det}(DL'-\lambda
I)=\mbox{det}(D^{\frac{1}{2}}L'D^{\frac{1}{2}}-\lambda I)
\end{equation}
valid for any $\lambda$, where ``det'' denotes the determinant and
$I$ is the $N\times N$ identity matrix, we have that the spectrum
of eigenvalues of matrix $G$ is equal to the spectrum of a
symmetric matrix defined as
\begin{equation}
H=D^{\frac{1}{2}}L'D^{\frac{1}{2}}.
\end{equation}
As a result, the eigenvalues of matrix $G$ are all nonnegative
real and the smallest eigenvalue is always zero.

\section{Simulations}

In our coupling method, giving the parameter $\beta$, for
$\alpha>0$, nodes with large degrees have stronger influence, and
for $\alpha<0$, nodes that bear few edges are more influential.
Parameter $\beta$ is exploited to eliminate the discrepancies
between the coupling signals that each node receive: Given
$\alpha$, when $\beta=1$, each node receives the equal quantum of
signals, when $\beta<1$, nodes that have larger sum of neighbors'
degrees are influenced more strongly, and when $\beta>1$, the
contrary situation occurs.

\begin{figure}
\scalebox{0.65}[0.65]{\includegraphics{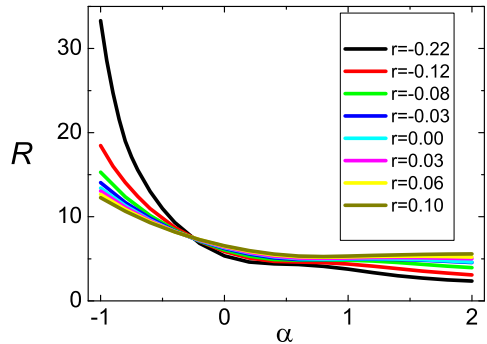}} \caption{(color
online) The eigenratio $R$ vs the parameter $\alpha$ at
$\beta=1.0$ for the generalized BA networks with different
assortative coefficients $r$. In all cases, the average degree is
$\bar{k}=6$, and the network size is $N=1024$. The data are
obtained over 10 realizations of network configurations.}
\end{figure}

\begin{figure}
\scalebox{0.65}[0.65]{\includegraphics{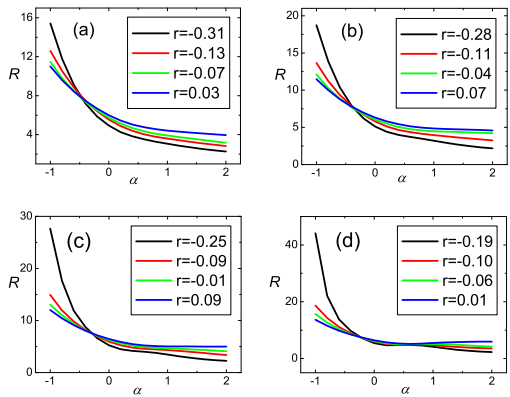}} \caption{(color
online) Ratio $R$ vs the parameter $\alpha$ at $\beta=1.0$ for the
generalized BA networks with different assortative coefficients
$r$. In each plot, the average degree is $\bar{k}=6$, and the
network size is $N=128$ (a), $N=256$ (b), $N=512$ (b), $N=2048$
(d). The data are obtained over 10 realizations of network
configurations.}
\end{figure}

Figure 1(a) shows the numerical values of eigenratio $R$ on the
parameter space ($\alpha$, $\beta$) for the well-known
Barab\'{a}si-Albert (BA) networks \cite{SFN}. To clearly exhibit
the effects of $\alpha$ and $\beta$ on $R$, we report the values
of $R$ as a function of $\alpha$ for different $\beta$ in figure
1(b). No matter what value the parameter $\beta$ takes, there
exists a region of $\alpha$, in which the eigenratio $R$ is
smaller than that of the case $\alpha=0$. That is to say, when
proper parameters are chosen, our coupling method can be even
better than the MZK method. Similar to the results obtained from
MZK method, $\beta=1.0$ corresponds to the optimal case (i.e. the
highest synchronizability). Hereinafter, we concentrate on the
case of $\beta=1$.

Note that, in the limit $\alpha=+\infty$ ($-\infty$), each node is
only influenced by the neighbor having the largest (smallest)
degree. The similar situation as mentioned in Ref. \cite{Chavez05}
appears: The original network approaches to a new configuration
that is connected by some effective directed edges \cite{ex1}, and
the new network, named the \emph{equivalent network}, may be
either connected or disconnected. In the disconnected case, the
eigenratio $R$ will approach to infinite, while in the connected
case, the eigenratio equals to 2 or some other larger constants.
Figure 2 illuminates three simple equivalent networks with
$\alpha=+\infty$; the former two are connected, and the third one
is disconnected. Their eigenratio are 2, 6.8284 and +$\infty$,
respectively. Figure 3 shows the changes of eigenratio $R$ with
the parameter $\alpha$ with $\beta=1$ of different network
configurations. When $\alpha>0.4$, the eigenratios for different
configurations go apart: Some approach 2 or other constants not
much larger than 2, while some go to infinity, due to whether the
equivalent networks are connected or not. In addition, from the
simulation, we find that with the increasing of network size, the
proportion of networks being disconnected when $\alpha=\infty$
($-\infty$) will increase sharply.

\begin{figure}
\scalebox{0.65}[0.65]{\includegraphics{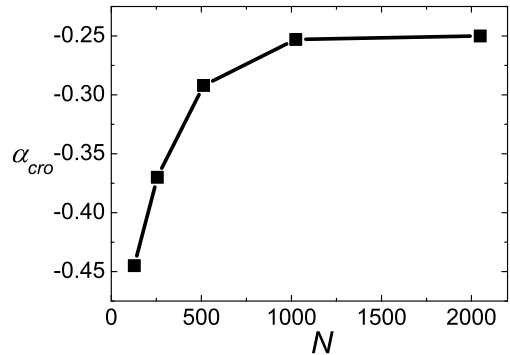}} \caption{The
crossed value of $\alpha$ vs the network size $N$ at $\beta=1.0$
for the generalized BA networks. In all cases, the average degree
is $\bar{k}=6$.}
\end{figure}

Next, we investigate the effects of degree-degree correlation on
the network synchronizability \cite{Bernardo2005}. The correlated
networks are generated by an extended BA algorithm
\cite{Dorogovtsev00,Krapivsky01}: Starting from $m_0$ fully
connected nodes, then, at each time step, a new node is added to
the network and $m$ ($<m_0$) previously existing nodes are chosen
to be connected to it with probability
\begin{equation}
p_i \propto \frac{k_i+k_0}{\sum_j(k_j+k_0)}
\end{equation}
where $p_i$ and $k_i$ denote the choosing probability and degree
of node $i$, respectively. By varying the free parameter $k_0$
$(>-m)$, one can obtain the scale-free networks with different
assortative coefficients $r$ \cite{Newman02,Newman03}.

Fig. 4 shows the relationship between eigenratio $R$ and the
parameter $\alpha$ for different assortative coefficients given
$\beta=1$. Interestingly, there exists a unique cross point at
$\alpha_{cro}\approx-0.25$. When $\alpha<\alpha_{cro}$, the
stronger assortative of network predicts better synchronizability,
while when $\alpha>\alpha_{cro}$, contrary phenomenon appears. In
Fig. 5, we report the simulation results for networks with
different sizes, which heightens the reliability of the existence
of this crossed behavior. Fig. 6 exhibits the cross point
$\alpha_{cro}$ as a function of the network size $N$: It increases
when $N$ is small, and will get steadily at about 0.25 for
sufficiently large $N$. Although it is interesting, unfortunately,
we are not able to provide a theoretical explanation about this
phenomenon.

\section{Conclusion and Discussion}
The hub nodes of a highly heterogeneous network always play the
major roles in determining the dynamical behaviors of the network.
In synchronizing process, the hub nodes simultaneously have two
effects. On the one hand, the throughput of these hub nodes are
too heavy, thus they will hinder the coupling signals'
transmission. On the other hand, they have great controlling
capability for their large number of coupling neighbors. The MZK
coupling method has taken into account the former point, and can
predict much better synchronizability than the uniform coupling
method. The present method further considers both aspects,
performs even better than MZK method, and shows that
synchronizability is maximum for a set of parameters that is
equivalent to the method introduced in Ref. \cite{MotterAIP05}.

Some previous works \cite{Yan2006,Yin2006,WangWX06,ZhouT06}
suggested that there exists some essentially common features
between network traffic and synchronization on a dynamical level
since the performance of them both are mainly determined by the
maximal betweenness, and many methods used to enhance the network
synchronizability can also improve the traffic conditions.
Therefore, a natural question raises: will the network throughput
increase if each node tends to receive information packets from
the large-degree neighbors?

\begin{acknowledgments}
This work is funded by the National Basic Research Program of
China (973 Program No.2006CB705500), the National Natural Science
Foundation of China (Grant Nos. 10472116, 10532060£¬70471033,
10547004 and 70571074), by the Special Research Funds for
Theoretical Physics Frontier Problems (No. A0524701), the
President Funding of Chinese Academy of Science, and the Graduate
Student Foundation of University of Science and Technology of
China under Grant Nos. KD2004008 and KD2005007.
\end{acknowledgments}


\begin{thebibliography} {Albert2000}
\bibitem{ex1} In this extreme case, only the directed edges containing nonzero coupling signals are effective edges, and can do contribution to the synchronizing process.
\end{thebibliography}
\end{document}